\RequirePackage{fix-cm}
\documentclass[twocolumn]{svjour3}          % twocolumn
\smartqed  % flush right qed marks, e.g. at end of proof
\usepackage{cite}
\usepackage{amsmath,amssymb,amsfonts}
\usepackage{algorithmic}
\usepackage{graphicx}
\usepackage{textcomp}

\usepackage{array}
\newcolumntype{C}[1]{>{\centering\arraybackslash} m{#1}}

\usepackage{tabularx}
\usepackage[table]{xcolor}% http://ctan.org/pkg/xcolor
\usepackage{epsfig}
\definecolor{forestgreen}{HTML}{228B22}
\definecolor{indigo}{HTML}{4B0082}

\usepackage{multirow}
\usepackage{eurosym}

\begin{document}

\title{Supervised Learning of the Global Risk Network Activation from Media Event Reports}
\titlerunning{Supervised Learning of the Global Risk Network Activation}
\author{Xiang Niu \and Gyorgy Korniss \and  Boleslaw K. Szymanski}

\authorrunning{Niu, Korniss, Szymanski} % if too long for running head

\institute{Xiang Niu \at
              Department of Computer Science \& Network Science and Technology Center, Rensselaer Polytechnic Institute (RPI), Troy, NY, USA \\
              \email{niux2@rpi.edu}           %  \\
           \and
           Gyorgy Korniss \at
              Department of Physics \& Network Science and Technology Center, Rensselaer Polytechnic Institute (RPI), Troy, NY, USA \\
              \email{korniss@rpi.edu}
           \and
           Boleslaw K. Szymanski \at
              Department of Computer Science \& Network Science and Technology Center, Rensselaer Polytechnic Institute (RPI), Troy, NY, USA \\
              \email{szymab@rpi.edu}
}

\date{}%Received: date / Accepted: date}

\maketitle

\begin{abstract}
The World Economic Forum (WEF) publishes annual reports of global risks which have the high impact on the world's economy. Currently, many researchers analyze the modeling and evolution of risks. However, few studies focus on validation of the global risk networks published by the WEF. In this paper, we first create a risk knowledge graph from the annotated risk events crawled from the Wikipedia. Then, we compare the relational dependencies of risks in the WEF and Wikipedia networks, and find that they share over 50\% of their edges. Moreover, the edges unique to each network signify the different perspectives of the experts and the public on global risks. To reduce the cost of manual annotation of events triggering risk activation, we build an auto-detection tool which filters out over 80\% media reported events unrelated to the global risks. In the process of filtering, our tool also continuously learns keywords relevant to global risks from the event sentences. Using locations of events extracted from the risk knowledge graph, we find characteristics of geographical distributions of the categories of global risks.

\keywords{Global risk network \and Knowledge graph \and Auto-detection \and Supervised learning}

\end{abstract}

\section{Introduction}

The World Economic Forum publishes annually the list of global risks and their properties. . The global risks are clustered into five categories: Economic, Environmental, Geopolitical, Societal and Technological~\cite{WEF2016}. These risks affect many aspects of lives of countless people. Economic risks may inflict tremendous damage on individuals, financial institutions, and governments. For example, in 2008, the asset bubble in several countries caused most of the major economies to undergo a recession lasting several years~\cite{erkens2012corporate}. Environmental risks affect both individuals and infrastructure but their impact is global when thousands of people die or are displaced by hurricanes or earthquakes~\cite{pielke2008normalized}. Geopolitical risks are mainly caused by a failure of national or global governance but may result in millions of people suffering from war~\cite{lee2003explaining}, hunger~\cite{sanchez2002soil}, and poverty~\cite{pogge2005world}. Technological risks are real and often unpredictable consequences of advanced technologies such as cyberattacks~\cite{rid2015attributing} and data leaks~\cite{harding2016panama} threatening well-being of individual and corporations. In short, all risks have critical ability to disrupt civilization at diverse levels, ranging from individuals to nations, so there is an is urgent need to develop tools for their discovery, analysis and prediction. 

The big challenge to progress in this direction is to identify activation of global risks from massive event records in Web media.
Therefore, risk-event labeling is a fundamental and necessary preparatory step before any in-depth study of risks, such as risk modeling~\cite{lin2017limits} and control~\cite{baloi2003modelling}, can be attempted. A high-quality dataset combined with the validated models are a foundation for global risks reliable discovery, analysis and prediction. The first systematic approach for such data collection effort was done and described in~\cite{szymanski2015failure}, in which the researchers manually collected monthly data points for each risk using online event resources such as articles, news, and Wikipedia to annotate events relevant to any global risk activation. The covered period ranged from
beginning of year 2000 to the end of year 2016. Considering the inaccuracy and incompleteness of manually labeled risk event records, researchers further improve these aspects of collected data by providing a detailed reason for selecting each risk event~\cite{niu2017evolution, niu2018evolution}. However currently, manual annotation is the only way to label risk events because of the difficulty to automate this operation caused by the unstructured format of event description. Here, we first propose a human-machine interaction model to augment the manual annotation process by precisely filtering out a large number of unrelated event. Then, we enhance this process by continuously expanding keyword sets for events using the supervised online-learning.

The auto-detection tool helps analysts to detect every single risk from event descriptions in Web media.
However, individual risk analysis is not sufficient because rarely a risk activation happens in isolation. According to the annual report from the World Economic Forum, the connectivity among risks is common and complicated. The global risks are clustered into five categories but their links includes both intra-category and inter-categories edges, both types of which are common. Most of the societal risks can activate under influence of economic (profound political and social instability), environmental (rapid and massive spread of infectious diseases) or geopolitical (large-scale involuntary migration) risks. The yearly WEF Global Risk Report is based on the experts' answers to the annual questionnaires that ask about each risk probability of activation, such activation costs to global economy and interconnectivity between risks. This dependence on expert opinions raises one important question: ``how well do the experts estimate risk probability of activation, impact, and interconnectivity?" followed by another: ``are their networks reliable, integrated, and representative?" To partially answer these questions, here, we create another risk network based only on the risk events discovered in Web media data. Then, we compare the resulting network with the WEF global risk network for the same risk nodes to assess both networks correctness and authoritativeness. We expect that the network from event risk data will be biased towards risks with high level of public awareness, which tend to be directly related to human lives, such as "{\it Asset bubble}" and "{\it Extreme weather event}".

Although the global risks attempt to represent events with impact on global economy and systematic impact on the entire world, most of the time, the risk events occur locally and have regional effects and influence. Thanks to the specific characteristic of the risk-events dataset, we extract local information from each event record and build connections between the locations and related risks. Having collected data for the period from year 2000 to 2014, we illustrate the different regional patterns of five risk categories on the world political maps.

\section{Risk detection}
\begin{figure}
    \centering
    \includegraphics[width=0.5\textwidth]{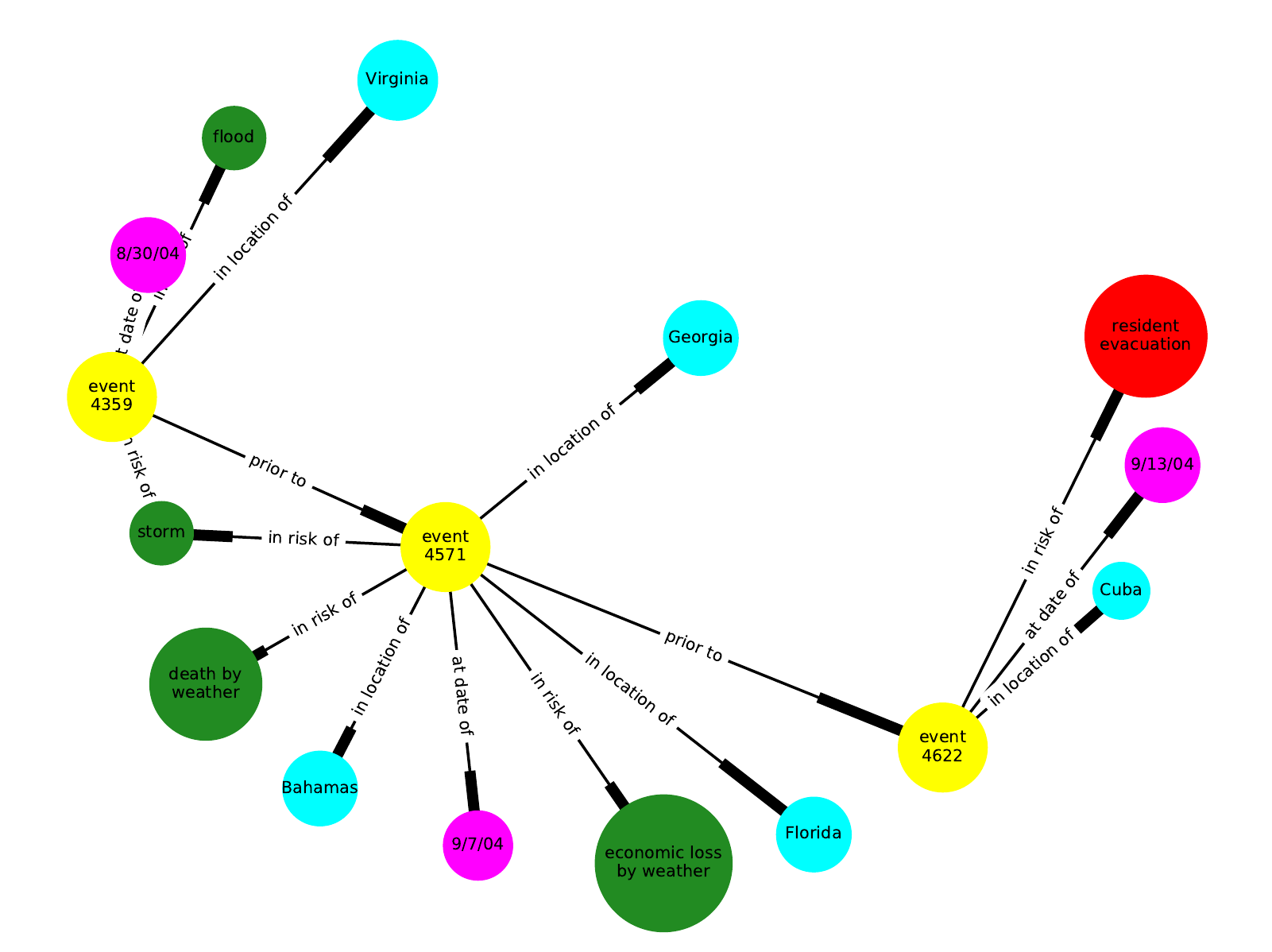}
    %\sidecaption
    \caption{An example of risk knowledge graph consisting of three events. Event 4359: "{\it Tropical Storm Gaston douses Richmond, Virginia, with up to 14 inches of rain, causing widespread flooding. (8/30/04)}". Event 4571: "{\it At least nine deaths in Florida, two deaths in the Bahamas, and one death in Georgia are blamed on the storm. Damage estimates range widely from US\$2 to US\$15 billion. (9/7/04)}". Event 4622: "{\it The Cuban government evacuates between 800,000 and 1.3 million people from coastal cities and developed areas. (9/13/04)}".
}
    \label{fig_knowleageGraphExample}
\end{figure}

In~\cite{szymanski2015failure}, authors investigate multiple event resources including articles, news, Wikipedia. Here, we only focus on the Wikipedia Current Events Portal (WCEP)~\cite{wikieventportal}, which comprehensively includes daily summaries of news events edited by crowdsourcing. Because of its popularity and completeness, the WCEP has been fully analyzed by comparing it with content written by professional journalists in WikiTimes project~\cite{tran2014indexing}. The conclusion of this project is that the WCEP is a representative source of public news thanks to its stable contribution volumes and crowdsourcing. The WikiTimes~\cite{tran2014indexing} also provides specific crawling methods and extracted dataset with over 50k events from the WCEP from the year 2000 to 2014, which are sufficient enough for the risk-event labeling and further studies. In the dataset, each event contains information detailed in Table~\ref{table_wiki_portal}. Fig.~\ref{fig_knowleageGraphExample} shows sample subnetwork of events. The event 4359, with description "{\it Tropical Storm Gaston douses Richmond, Virginia, with up to 14 inches of rain, causing widespread flooding}" and entities "{\it Tropical Storm Gaston}", "{\it Virginia}" and "{\it Mark Warner}", are parts of story "{\it 2004 Atlantic hurricane season}" occurring on Aug. 30, 2004, reported by ABC-NEWS. The event 4571 is described as "{\it At least nine deaths in Florida, two deaths in the Bahamas, and one death in Georgia are blamed on the storm. Damage estimates range widely from US\$2 to US\$15 billion}". The event 4622 reports "{\it The Cuban government evacuates between 800,000 and 1.3 million people from coastal cities and developed areas}". The last two events are parts of the same story about "{\it 2004 Atlantic hurricane season}". 
\begin{table}
\centering
\begin{tabular}{|c|c|} \hline
\textbf{Concept} &  \textbf{Description} \\ \hline
Event & sentences of a news summary \\ \hline
Story & a story in Wikipedia of which the event is part \\ \hline
Category & a category to which the event belongs \\ \hline
Date & the event date of occurrence \\ \hline
Entities & a list of entities in the event sentences \\ \hline
References & external sources for the same event \\ \hline
\end{tabular}
\caption{Structured event information extracted from the WCEP~\cite{tran2014indexing}.}
\label{table_wiki_portal}
\end{table}

\begin{figure*}
    \centering
    \includegraphics[width=\textwidth]{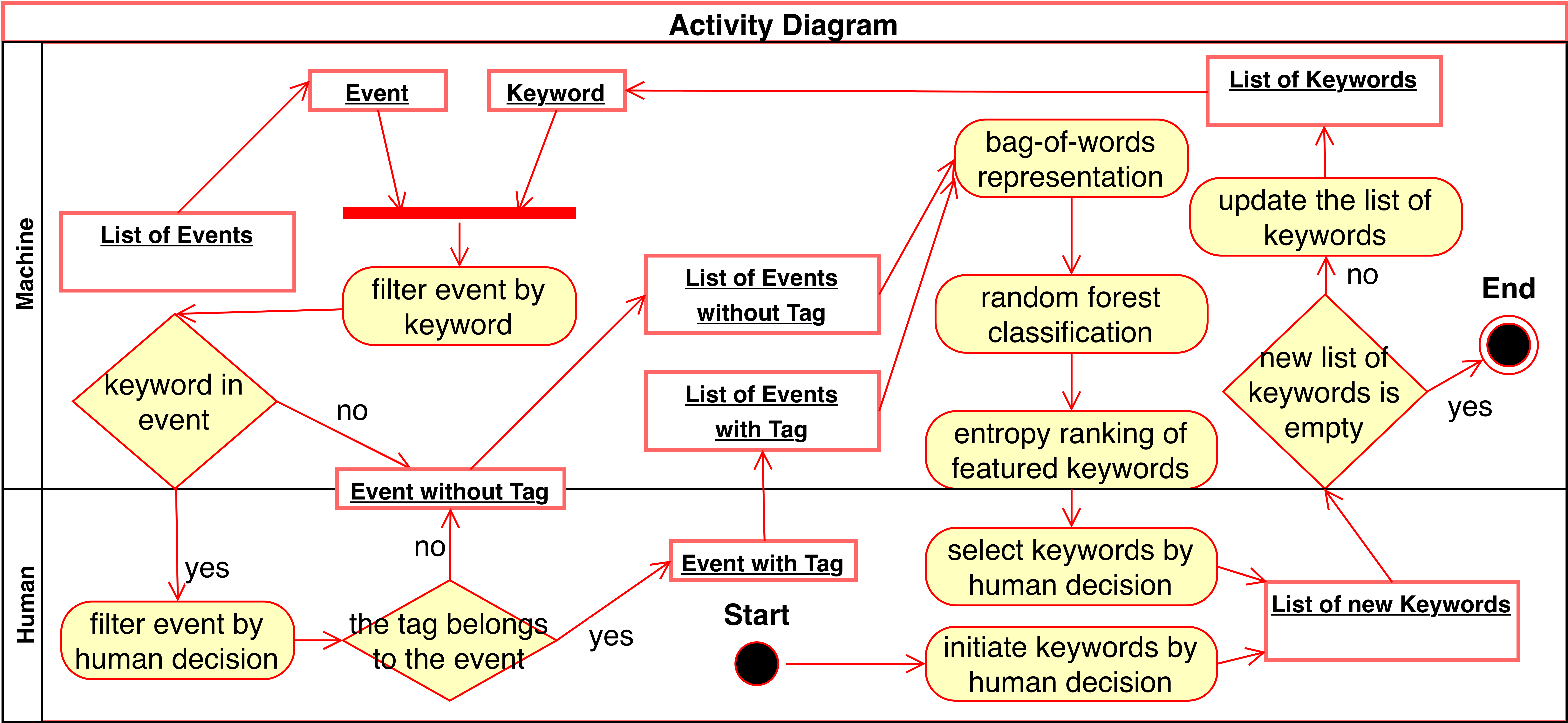}
    %\sidecaption
    \caption{The activity diagram of risk auto-detection module.}
    \label{risk_learning}
\end{figure*}

Manual annotation requires a person to go first through the event description word by word and select the risk related keywords, then label the event with risk tags according to the keywords. 
For example, for the event 4359, two keywords "{\it storm}" and "{\it flooding}" may be selected and then the event will be labeled with tags "{\it storm}" and "{\it flood}" accordingly. Both tags belong to the same risk "{\it 10. Extreme weather events}", as shown Table~\ref{table_risk_ids_2} in Appendix. The event 4571 contains two more tags "{\it death}" and "{\it economic loss}" in addition to "{\it storm}". Thus, this event shows the activity of one more risk "{\it 11. Failure of climate-change mitigation and adaptation}", see Table~\ref{table_risk_ids_2} in Appendix. The event 4622 contains only one tag "{\it evacuation}", which belongs to  risk "{\it 23. Large-scale involuntary migration}", as shown in Table~\ref{table_risk_ids_4} in Appendix. Thus, the three events represent the activities of three risks in two risk categories: environmental risks, as seen in Table~\ref{table_risk_ids_2} in Appendix, and societal risks according to Table~\ref{table_risk_ids_4} in Appendix.

With the development of modern information society, there is a large number of news events described all over the internet. Specifically, around three thousand new events are generated in the WCEP every year amounting to about 250 per each month and over fifty thousand events in the past 15 years. Given 29 risk labels and around 20 words per event description, an expert would have to go through $3000\times30\times20$ so nearly two million comparisons to process just one year of events data in the WCEP, which is a heavy workload. Besides, during the labeling, only a few (around two thousand) of the fifty thousand events are relevant to global risks. These observations motivated us to create an auto-detection tool to help analysts by filtering out the irrelevant events, thereby saving time in risk-event labeling. Our risk auto-detection module is depicted in Fig.~\ref{risk_learning}. The detailed processing proceeds according to the following steps:
\begin{itemize}
    \item
    {\bf Step 1}. Initially, an analyst prepares a few keywords and tags for each risk according to the description of 29 risks in Appendix based on ~\cite{WEF2016}.
    \item
    {\bf Step 2}. If the given list of new pairs with keyword $k$ and tag $t$ $\langle k, t \rangle$ is empty, processing terminates, otherwise, these pairs are merged into keyword dictionary.
    \item
    {\bf Step 3}. For each event $e$ and a list of keywords $K$, each paired with some tag $t$, if a keyword $k\in K$ appears in the event sentence manually accepted as relevant, the event $e$ is labeled with the tag $t$ as $\langle e, t^+ \rangle$, otherwise, the event $e$ is labeled as $\langle e, t^- \rangle$.
    \item
    {\bf Step 4}. We use a bag-of-words representation $b = \langle w_1, c_1 \rangle,$ $...,$ $\langle w_n, c_n \rangle$ for each event sentence, where $w_i$ is the $ith$ word of event sentence in $e$, $c_i$ is the count of appearances of the word $w_i$ for $n$ unique words. For each tag $t$, we have a list of events: $\langle e, t^+ \rangle$ is a positive instance with bag-of-words representation $b^{+}\in B^{+}$, while $\langle e, t^- \rangle$ is a negative instance with bag-of-words representation $b^{-}\in B^{-}$ in the event sentence. 
    \item
    {\bf Step 5}. With random forest classifier given training dataset $B^{+}$ and $B^{-}$, we get entropy ranking of featured keywords $w_1, w_2, ..., w_m$, where $m$ is the total number of unique words in $B^{+}$ and $B^{-}$. The words with high rank are important for the classification. 
    \item
    {\bf Step 6}. Based on the entropy ranking list, we select meaningful keywords from top $a$ (default $a=5$) new ones and create a list of $\langle k, t \rangle$. Then, processing restarts at step 2.
\end{itemize}
Tables~\ref{table_risk_ids_1}-\ref{table_risk_ids_5} in Appendix show the list of selected tags and keywords of 29 risks with current dataset up to year 2014. With new events data, the module can update the list further and improve the filtering accuracy. In step 3, the events without any keyword $k\in K$ or rejected by analysts are filtered out as non-relevant events. 

%Bag-of-words, Random forest, Keywords extraction

\begin{figure}
    \centering
    \includegraphics[width=0.5\textwidth]{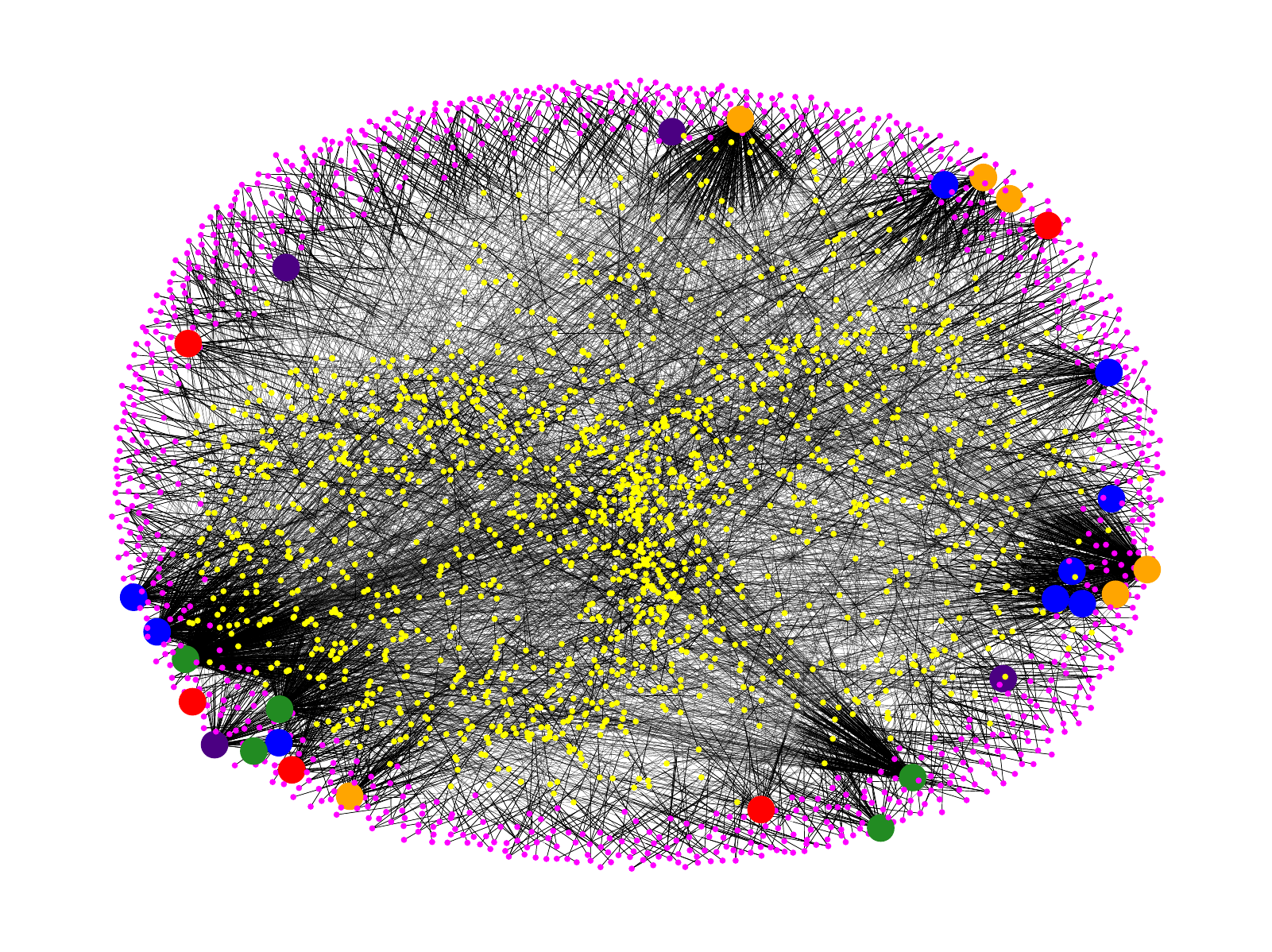}
    %\sidecaption
    \caption{The entire risk knowledge graph with small yellow nodes representing events, small purple nodes representing the date, and large nodes representing thirty risks from five colored categories: economic, environmental, geopolitical, societal, and technological.}
    \label{fig_knowleageGraphWhole}
\end{figure}

\begin{figure*}
    \centering
    \includegraphics[width=\textwidth]{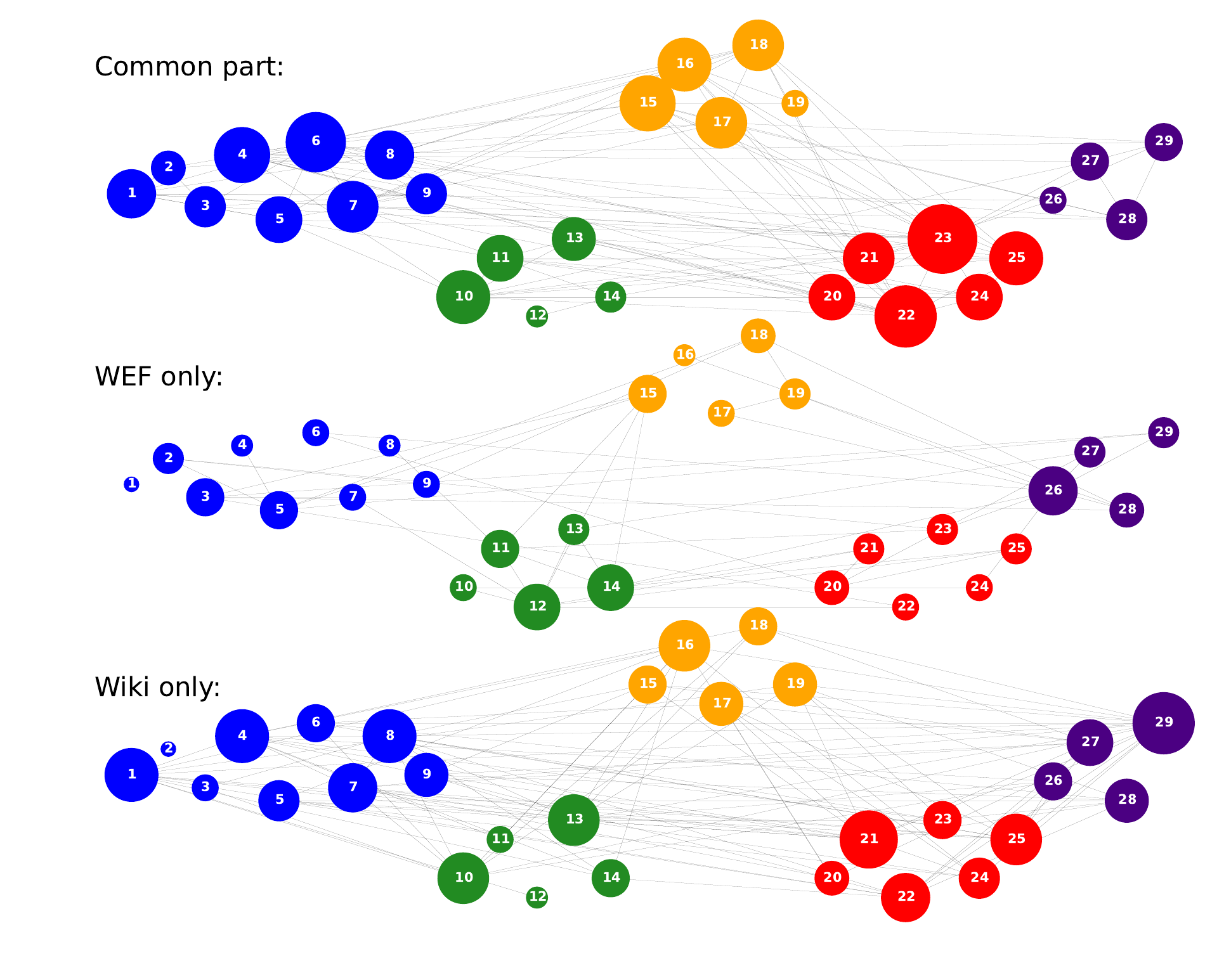}
    \caption{A comparison between two risk networks, one created by the WEF and the other built by us from Wiki events. The first subfigure is the common part extracted from the two networks, the second subfigure contains edges unique to the WEF network and the third one includes relational dependencies unique to the Wiki network.}
    \label{fig_network_cmp}
\end{figure*}

\section{Risk network}
The auto-detection module produces a structured event dataset containing rich independent information such as risk labels, stories, event descriptions, their occurrence locations, and dates. Based on that, with the specific principal (story) and subordinate (event) relationship, we further build event correlations, see Fig.~\ref{fig_knowleageGraphExample}. Any two events are considered related when they belong to the same story. The knowledge graph includes the following four types of relationships:
\begin{itemize}
    \item
    At date of: an event $e$ happened at date $d$. Any event $e$ has a unique $d$ associated with it. But one date $d$ may correspond to multiple events. 
    \item
    At location of: an event $e$ happened at location $l$ or at location relevant to $l$. Any event $e$ may happen (affect) multiple locations, e.g., Fig.~\ref{fig_knowleageGraphExample} event 4571 affected locations at Bahamas, Florida and Georgia. One location $l$ may also correspond to multiple events. 
    \item
    Prior to: an event $e_1$ happened prior to another event $e_2$ when they are in the same story and the date $d_1$ of $e_1$ precedes the date $d_2$ of $e_2$.
    \item
    In risk of: an event $e$ is in risk of $t$ after the event sentence is labeled by the tag $t$ by the auto-labeling module, see Fig.~\ref{risk_learning}.
\end{itemize}
Fig.~\ref{fig_knowleageGraphWhole} depicts the entire knowledge graph that contains all relationships between any two events, and between two pairs for event $e$, ($e$, risk) and ($e$, date). Here, we group all tags $t_1$, $t_2$, ... belonging to the same risk $r$ into one risk node. For example, in Fig.~\ref{fig_knowleageGraphExample}, event 4359 is in risk of "{\it storm}" and "{\it flood}", which correspond to the same risk "{\it 10. Extreme weather event}".

Additionally, we extracted the risk network (Wiki in Fig.~\ref{fig_network_cmp}) from the entire knowledge graph (Fig.~\ref{fig_knowleageGraphWhole}). For each pair of events ($e_1$, $e_2$), when $e_1$ is prior to $e_2$, $e_1$ is in risks of $R_1$ and $e_2$ in risks of $R_2$, we create an edge between each pair of ($r_{i}$, $r_{j}$), $r_i, r_j \in R_1 \bigcup R_2, i\ne j $. 
For example in Fig.~\ref{fig_knowleageGraphExample}, event 4359 is in risks of "{\it 10. Extreme weather event}" $r_{10}$, event 4571 is in risks of "{\it 10. Extreme weather event}" $r_{10}$ and "{\it 11. Failure of climate-change mitigation and adaptation}" $r_{11}$, event 4622 is in risks of "{\it 23. Large-scale involuntary migration}" $r_{23}$. Thus, we have three connections $r_{10}-r_{11}$, $r_{10}-r_{23}$ and $r_{11}-r_{23}$.
To make a comparison of so created network with the undirected and unweighted risk network created by the World Economic Forum (WEF) in report~\cite{WEF2016}, we transform the Wiki network to be also undirected and unweighted, as seen in Fig.~\ref{fig_network_cmp}, where the size of each risk node represents its degree. In general, the two networks are similar in nodes sizes and connections. There are 170 edges in WEF network and 224 edges in Wiki networks, with the number of common edges as high as 121, see Table~\ref{table_network_edge_cmp}. Specifically, the risk "{\it 23. Profound social instability}" has the largest number (18) of common related risks between the two networks. 

We also highlight the different risk dependencies in WEF and in Wiki networks in Fig.~\ref{fig_network_cmp}. For example, in Wiki network, risk "{\it 1. Asset bubbles in a major economy}" influences and is influenced also by risks "{\it 4. Failure/shortfall of critical infrastructure}", "{\it 22. Large-scale involuntary migration}", and etc. The degree differences between the two risk networks arise from the different views of the risks by experts and the public. The risks with a higher degree in Wiki network tend to be a focus of public concerns, such as "{\it 1. Asset bubbles in a major economy}", "{\it 10. Extreme weather events}", "{\it 17. Large-scale terrorist attacks}", "{\it 21. Food crises}" and "{\it 29. Massive incident of data fraud/theft}". Nevertheless, the risks with a higher degree in WEF network tend to be a focus of expert concerns, often novel or merely newly arising, such as "{\it 2. Deflation in a major economy}", "{\it 12. Major biodiversity loss and ecosystem collapse}", and "{\it 26. Adverse consequences of technological advances}". In addition, in Table~\ref{table_network_edge_cmp}, the common part of the two networks includes most of the intra-group edges. After extracting it from the WEF and the Wiki network, almost no intra-group edges left, especially in Wiki network. Moreover, in Table~\ref{table_network_degree_cmp}, after extracting the common part, the WEF network becomes fragmented with extremely low average clustering coefficients in each group and in the entire network.

\begin{table*}
\centering
\begin{tabular}{|c|c|c|c|c|c|c|c|c|c|c|c|c|} \hline
\textbf{Networks} & \multicolumn{2}{c|}{\textbf{Economic}}  &   \multicolumn{2}{c|}{\textbf{Environmental}} &  \multicolumn{2}{c|}{\textbf{Geopolitical}} &  \multicolumn{2}{c|}{\textbf{Societal}} &  \multicolumn{2}{c|}{\textbf{Technological}} & \textbf{Whole Network} \\ \hline
\textbf{} &  \textbf{Intra} &  \textbf{Inter}  &  \textbf{Intra} &  \textbf{Inter} &  \textbf{Intra} &  \textbf{Inter} &  \textbf{Intra} &  \textbf{Inter} &  \textbf{Intra} &  \textbf{Inter} & \textbf{Number of edges}  \\ \hline
Common part & 21 & 36 & 4 & 22 & 8 & 29 & 12 & 47 & 3 & 12 & 121 \\ \hline 
WEF only & 4 & 13 & 6 & 14 & 2 & 11 & 2 & 13 & 3 & 13 & 49 \\ \hline
WEF & 25 & 49 & 10 & 36 & 10 & 40 & 14 & 60 & 6 & 25 & 170 \\ \hline
Wiki only & \textbf{6} & 50 & \textbf{0} & 28 & \textbf{0} & 34 & \textbf{1} & 45 & \textbf{0} & 35 & 103 \\ \hline
Wiki & 27 & 86 & 4 & 50 & 8 & 63 & 13 & 92 & 3 & 47 & 224 \\ \hline
\end{tabular}
\caption{The number of intra-group and inter-group edges of each risk category and network. WEF represents the network published by the World Economic Forum; Wiki represents the network created by us from Wiki events. Common part represents the edges existing in both networks. WEF only represents the edges unique to the WEF network, while Wiki only includes the edges unique to the Wiki network.}
\label{table_network_edge_cmp}
\end{table*}

\begin{table*}
\centering
\begin{tabular}{|c|c|c|c|c|c|c|c|c|c|c|c|c|c|} \hline
\textbf{Networks} & \multicolumn{2}{c|}{\textbf{Economic}}  &   \multicolumn{2}{c|}{\textbf{Environmental}} &  \multicolumn{2}{c|}{\textbf{Geopolitical}} &  \multicolumn{2}{c|}{\textbf{Societal}} &  \multicolumn{2}{c|}{\textbf{Technological}} & \multicolumn{2}{c|}{\textbf{Whole Network}} \\ \hline
\textbf{} &  $\langle k \rangle$ &  $\langle c \rangle$  &  $\langle k \rangle$ &  $\langle c \rangle$ &  $\langle k \rangle$ &  $\langle c \rangle$ &  $\langle k \rangle$ &  $\langle c \rangle$ &  $\langle k \rangle$ &  $\langle c \rangle$ & $\langle k \rangle$ &  $\langle c \rangle$ \\ \hline
Common part & 8.67 & 0.6 & 6.0 & 0.49 & 9.0 & 0.69 & 11.83 & 0.59 & 4.5 & 0.47 & 8.34 & 0.58 \\ \hline
WEF only & 2.33 & \textbf{0.0} & 5.2 & \textbf{0.06} & 3.0 & \textbf{0.31} & 2.83 & \textbf{0.17} & 4.75 & \textbf{0.01} & 3.38 & \textbf{0.1} \\ \hline
WEF & 11.0 & 0.61 & 11.2 & 0.66 & 12.0 & 0.57 & 14.67 & 0.59 & 9.25 & 0.48 & 11.72 & 0.59 \\ \hline
Wiki only & 6.89 & 0.35 & 5.6 & 0.42 & 6.8 & 0.35 & 7.83 & 0.32 & 8.75 & 0.22 & 7.1 & 0.34 \\ \hline
Wiki & 15.56 & 0.74 & 11.6 & 0.81 & 15.8 & 0.81 & 19.67 & 0.72 & 13.25 & 0.85 & 15.45 & 0.78 \\ \hline
\end{tabular}
\caption{The average degree $\langle k \rangle$ and average clustering coefficient $\langle c \rangle$ of each risk category and network. WEF represents the network published by the World Economic Forum; Wiki represents the network created by us from Wiki events. Common part includes the edges existing in both networks. WEF only represents the edges unique to the WEF network, while Wiki only includes the edges unique to the Wiki network.}
\label{table_network_degree_cmp}
\end{table*}

\begin{figure}
    \centering
    \includegraphics[width=0.5\textwidth]{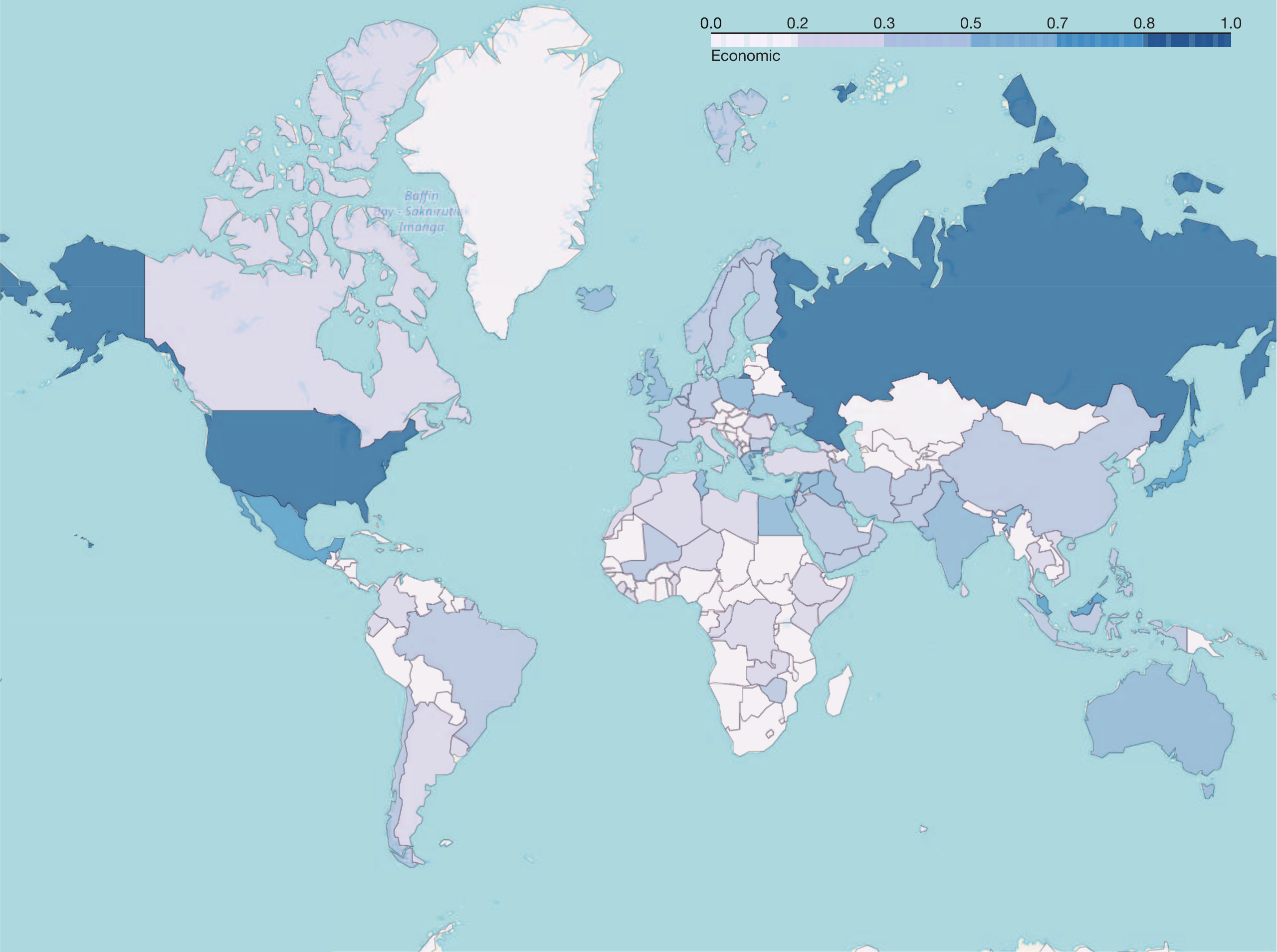}
    %\sidecaption
    \caption{Economic risks heat-map over political map of the world.}
    \label{fig_eco_map}
\end{figure}

\section{Risk activities heat maps layered over the political maps of the world}
\begin{figure}
    \centering
    \includegraphics[width=0.5\textwidth]{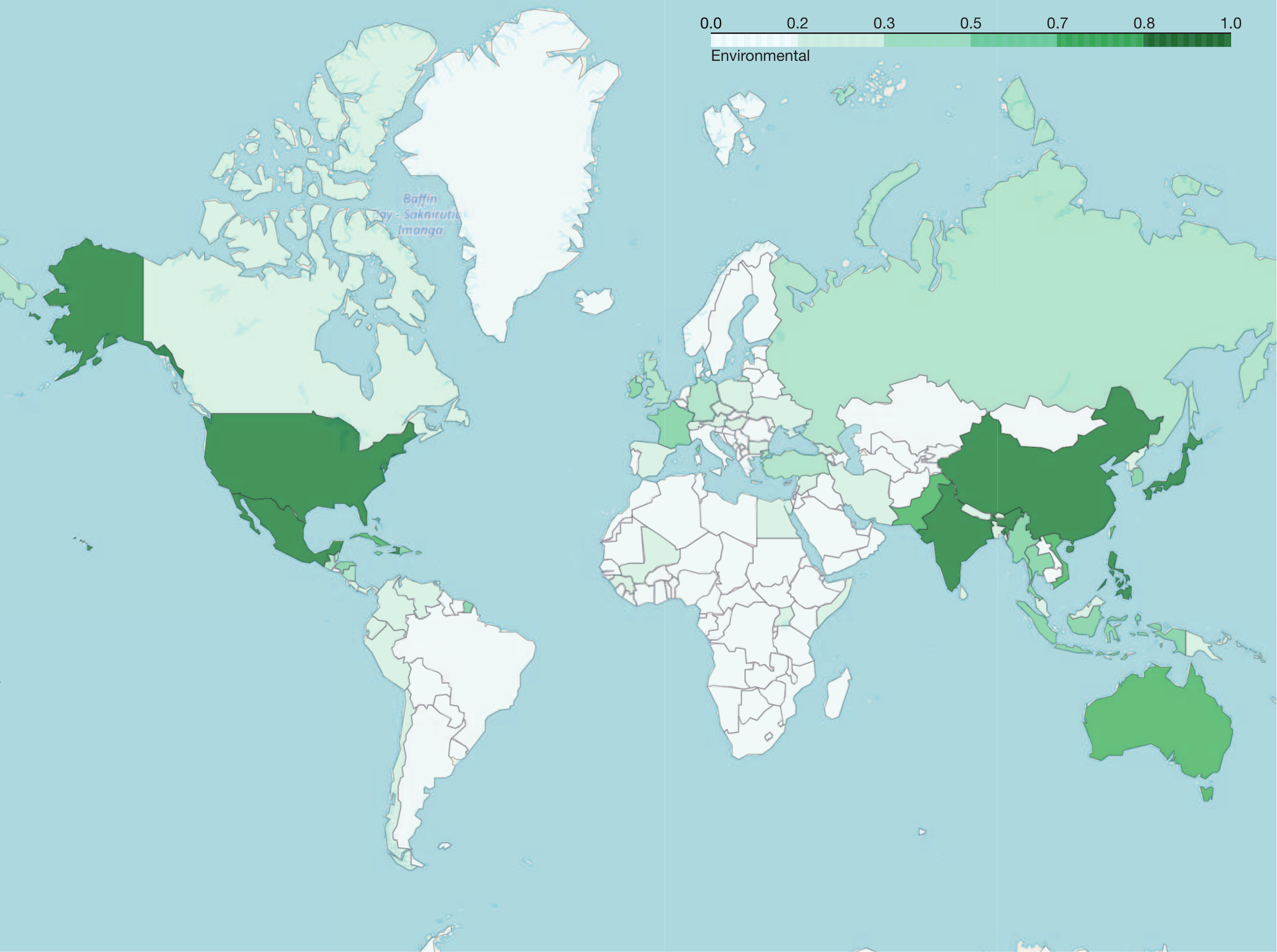}
    \caption{Environmental risks heat-map over political map of the world.}
    \label{fig_env_map}
\end{figure}

\begin{figure}
    \centering
    \includegraphics[width=0.5\textwidth]{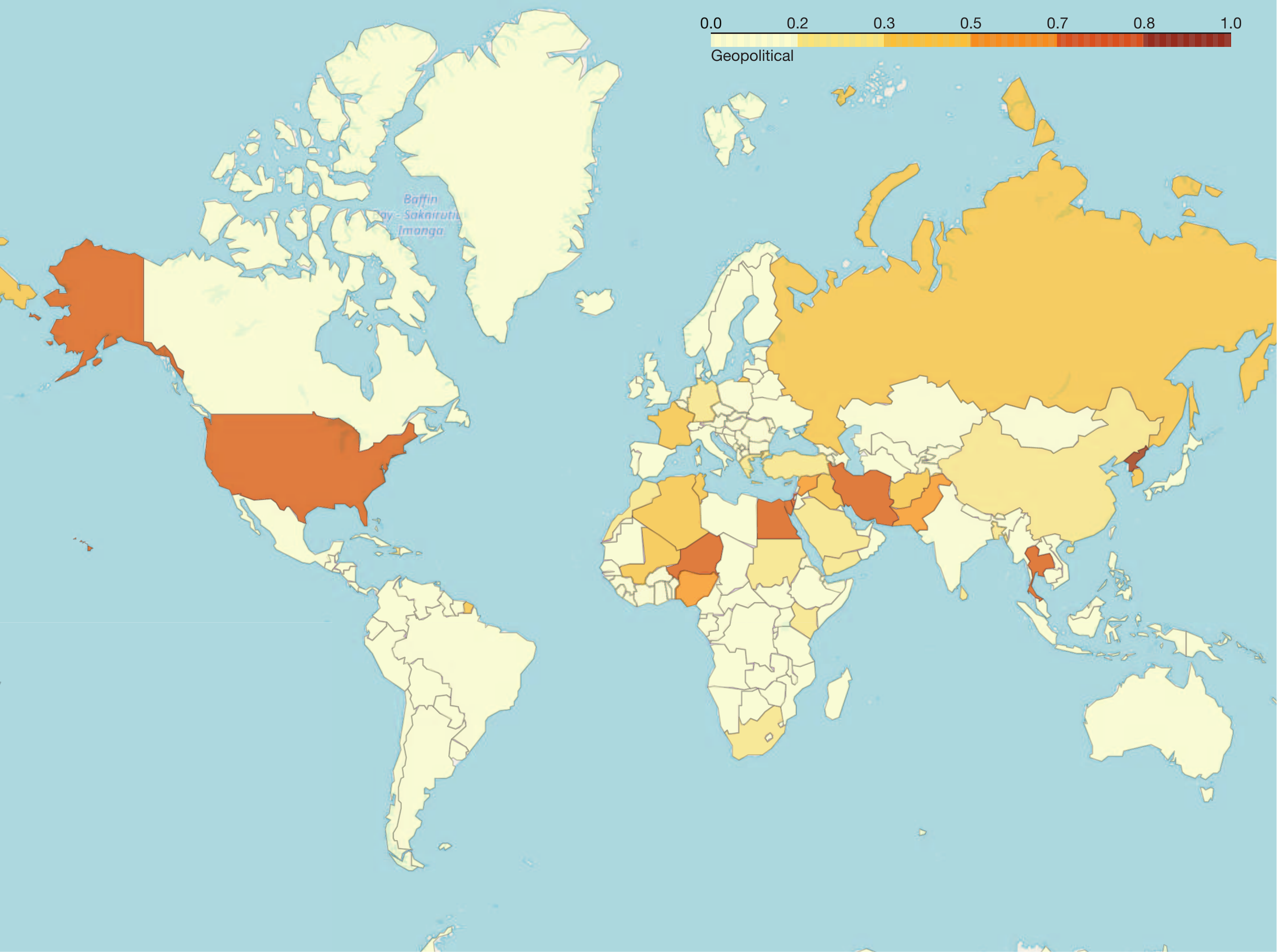}
    \caption{Geopolitical risks heat-map over political map of the world.}
    \label{fig_geo_map}
\end{figure}

\begin{figure}
    \centering
    \includegraphics[width=0.5\textwidth]{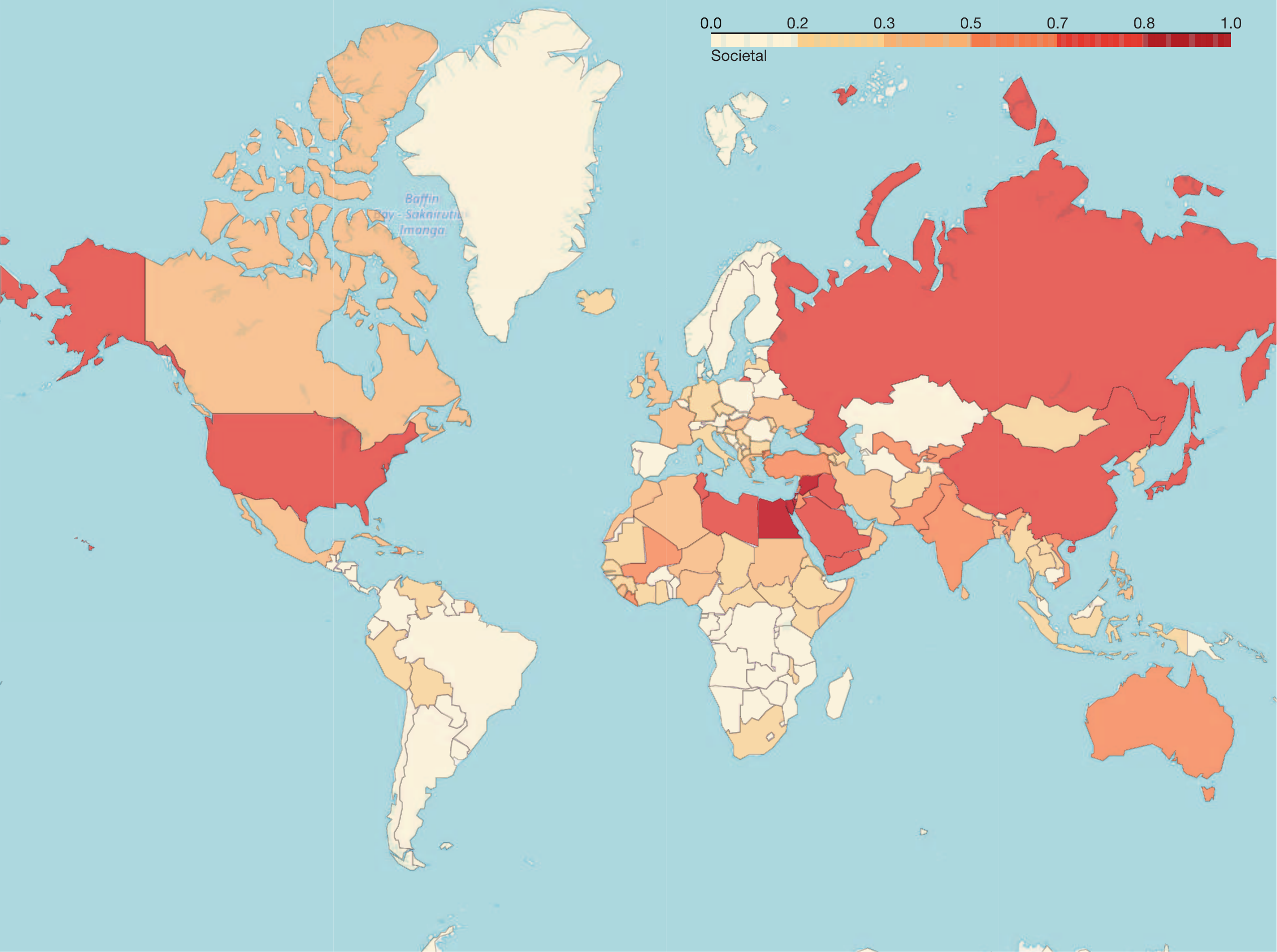}
    \caption{Societal risks heat-map over political map of the world.}
    \label{fig_soc_map}
\end{figure}

\begin{figure}
    \centering
    \includegraphics[width=0.5\textwidth]{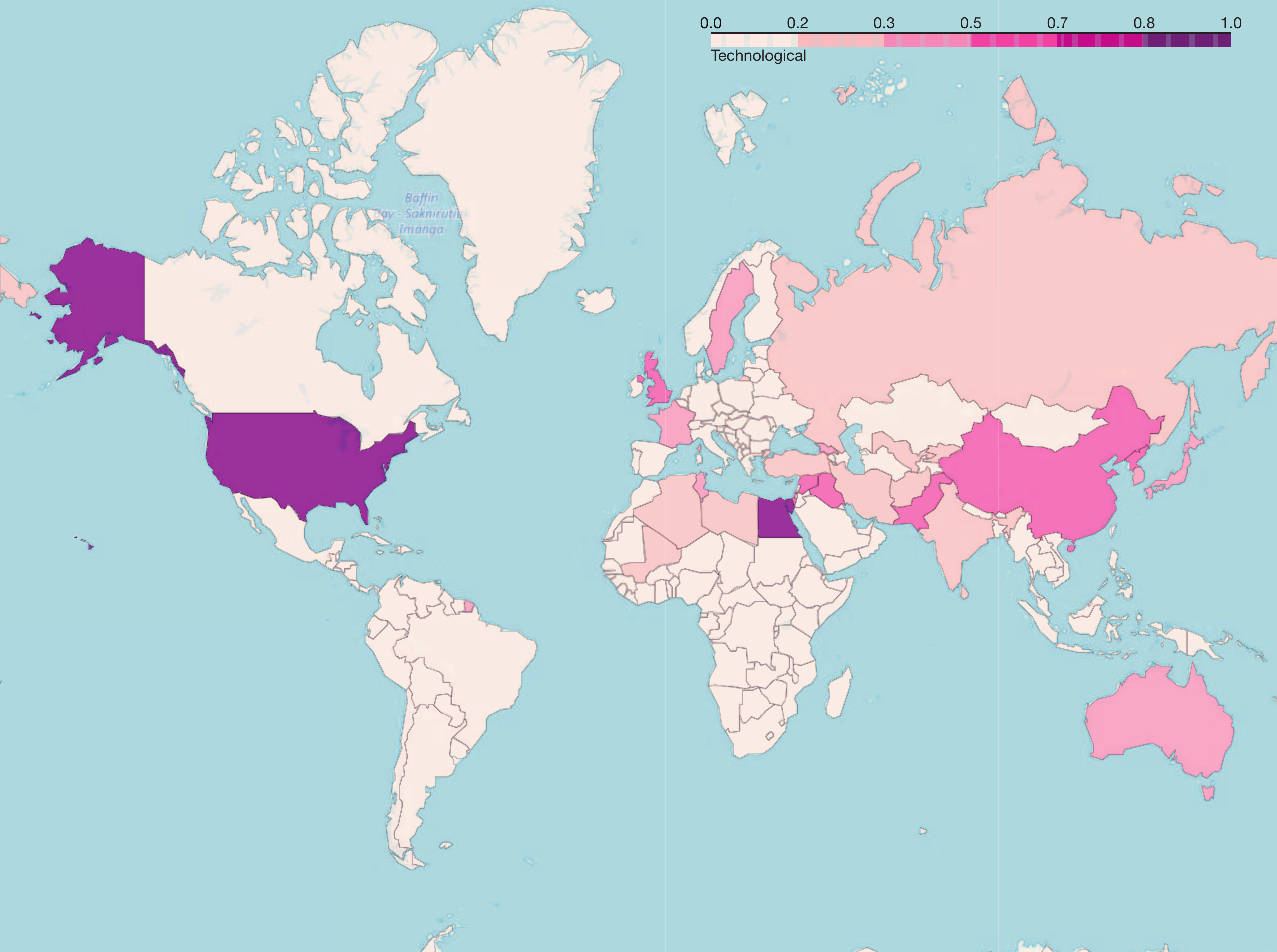}
    \caption{Technological risks heat-map over political map of the world.}
    \label{fig_tec_map}
\end{figure}

Using the risk knowledge graph, we extract connections between locations and risks. For any event $e$ in risk $r\in R$ that occurred at location $l\in L$, where $R$ is one of five risk categories, $L$ is a country to which location $l$ belongs, we record an occurrence of pair $(R, L)$. After processing all events, we obtain the number of occurrences of each pair $(R, L)$. For each risk category $R$, we normalize the number of occurrences in all countries $L$ by taking logarithm of a ratio of number of occurrence+1 in the given country to 1+the maximum number of occurrences in any country for category $R$. We show the results as a heat-maps layered over the political maps of the world shown in Figs.~\ref{fig_eco_map}-\ref{fig_tec_map}. Each of these figures displays normalized levels of activities of one of the five risk categories. For example, in economic risks, the economic risk-events related to Russia occur 42 times, related to Japan 16 times, to India 6 times, to Italy 2 times. These scores after the normalization become 1, 0.75, 0.52, 0.29, respectively. 

Most of the economic risks happen in the developed countries or countries with high volume of trading resources, industrial products or services, such as Russia (1.00), U.S. (0.95), Malaysia (0.78), Japan (0.75), Mexico (0.72), Cyprus (0.68), Australia (0.61), Greece (0.61). The most impactful story in Russia is "{\it 2008 Russian financial crisis}", in the U.S. "{\it Subprime mortgage crisis}", in Malaysia "{\it Malaysia Airlines Flight 370}", in Japan "{\it Automotive industry crisis of 2008-10}", in Mexico "{\it Mexican Drug War}", in Cyprus "{\it 2012-13 Cypriot financial crisis}", in Australia "{\it August 2011 stock markets fall}", and in Greece "{\it Greek government-debt crisis}". Those economic risks are related to a failure of financial mechanisms such as stock (Russia, Australia), mortgage (U.S.), banking (Cyprus, Greece), illicit trade, such as smuggling (Mexico), and decreased demand for manufacturing products, such as cars (Japan).

The environmental risks usually happen in coastal countries, especially those near Atlantic, Pacific, or Indian Oceans, such as U.S. (1.00), Philippines (1.00), China (0.97), Japan (0.91), India (0.91), Mexico (0.90), Haiti (0.87), Pakistan (0.83), Cuba (0.71).
"{\it Atlantic hurricane season}" mainly affects the U.S., Mexico, Haiti, Cuba.
"{\it Pacific typhoon season}" mainly affects the Philippines, China, Japan.
"{\it North Indian Ocean cyclone season}" mainly affects India.
Beyond that, some other disasters also happen in the U.S. "{\it 2011 Mississippi River floods}", in China "{\it 2008 Sichuan earthquake}", in Japan "{\it 2011 Tōhoku earthquake and tsunami}", in India "{\it 2004 Indian Ocean earthquake and tsunami}", in Mexico "{\it Deepwater Horizon oil spill}", and in Pakistan "{\it 2010 Pakistan floods}".

The geopolitical risks are most frequent in countries having unstable regime or involved in international conflicts, such as North Korea (1.00), Thailand (0.80), Iran (0.80), U.S. (0.72), Egypt (0.72), Pakistan (0.62), Syria (0.62), Nigeria (0.62), Iraq (0.38). 
Risk "{\it 16. Interstate conflict with regional consequences}" is between the U.S. and Iraq in "{\it Iraq War}".
Risk "{\it 17. Large-scale terrorist attacks}" is in Pakistan "{\it Terrorism in Pakistan}" and Nigeria "{\it Islamist insurgency in Nigeria}".
Risk "{\it 18. State collapse or crisis}" is in Thailand "{\it Thai coup d'état}", Egypt "{\it Egyptian coup d'état}", and Syria "{\it Syrian civil war}".
Risk "{\it 19. Weapons of mass destruction}" is in North Korean "{\it North Korean nuclear test}" and Iran "{\it Nuclear program of Iran}".

%newziland

Societal risks, are often a consequence of geopolitical risks, and usually occur in the same countries as the latter group of risks does, such as Egypt (1.00), Syria (0.98), Yemen (0.80), Libya (0.77), Tunisia (0.77), China (0.76), Russia (0.76), Saudi Arabia (0.73). The most influential story in Egypt is "{\it Egyptian Revolution of 2011}", in Syria "{\it Syrian civil war}", in Yemen "{\it Yemeni Revolution}", in Libya "{\it Libyan civil war}", in Tunisia "{\it Tunisian Revolution}", in China "{\it 2009 Ürümqi riots}", in Russia "{\it Snow Revolution}", and in Saudi Arabia "{\it Arab Spring}".

Technological risks frequently happen either in a scientifically developed country such as U.S. (1.00) and U.K. (0.51) or in an unstable nations such as Egypt (0.85), Syria (0.66), Iraq (0.51), Pakistan (0.51), North Korea (0.51). 
When information technology is highly developed, risk "{\it 29. Massive incident of data fraud/theft}" is often active because of a massive amount of official and personal data online, such as "{\it United States diplomatic cables leak}" in the U.S. and "{\it News International phone hacking scandal}" in the U.K. 
Related geopolitical risks may also cause other technological risks. For example, in Egypt, a long-lasting internet blackout and censorship was triggered by "{\it Egyptian Revolution of 2011}"; in North Korea, risk "{\it 28. Large-scale cyberattacks}" was caused by "{\it 2013 Korean crisis}".

\section{Conclusion}
Our paper contributes the first event-driven analysis of global risks using the complete Wikipedia Event Portal dataset over the years 2000 to 2014. To overcome the limitation of risk-event manual annotation used in the current research~\cite{szymanski2015failure, niu2018evolution}, here we introduce a novel risk auto-detection module, which filters out over 80\% of data about the non-relevant events and reduces significantly volume of data to be manually labeled. This approach can be used in other machine learning scenarios when the true negative instances account for a large proportion of the dataset during the supervised learning classification. With the help from auto-detection tool and sufficient effort devoted to manual labeling, we created the complete and structured dataset from Wikipedia Event Portal, which can be studied in-depth in the future. This data set contains two thousand labeled risk events extracted from over 50 thousand events. These labeled risks are structured hierarchically with events linked to stories, locations, dates, categories, and references creating a knowledge network.

From this rich dataset, we build a relational network, which is surprisingly similar to the network created by the WEF experts~\cite{WEF2016}.
Moreover, the differences between the two networks reflect the differences in risk views of the expert and the public. The experts show concerns about yet unseen risks that may arise in the future or have an enormous impact on all humanity, while the public tends to focus on risks that matter more to everyday life. By studying the local effects of risks, we find that the economic risks arise in countries with high volume of trade. The environmental risks are common in coastal areas. The geopolitical risks are concentrated mainly in the Middle East. Most of the societal risks are the consequences of geopolitical risks and thus occur in the same countries as the former. The technological risks arise in scientifically developed countries and in the countries in conflict with other nations.

\section*{Acknowledgment}
This work was supported in part by the Army Research Office under grant W911NF-16-1-0524 and by the Army Research Laboratory under Cooperative Agreement Number W911NF-09-2-0053 (NS CTA). The views and conclusions contained in this document are those of the authors and should not be interpreted as representing the official policies either expressed or implied of the Army Research Laboratory or the U.S. Government.

\bibliographystyle{spphys}
\bibliography{reference}

\begin{thebibliography}{10}
\providecommand{\url}[1]{{#1}}
\providecommand{\urlprefix}{URL }
\expandafter\ifx\csname urlstyle\endcsname\relax
  \providecommand{\doi}[1]{DOI \discretionary{}{}{}#1}\else
  \providecommand{\doi}{DOI \discretionary{}{}{}\begingroup
  \urlstyle{rm}\Url}\fi

\bibitem{WEF2016}
{W}orld {E}conomic {F}orum {G}lobal {R}isks {R}eport. (2016).
\newblock
  \urlprefix\url{https://www.weforum.org/reports/the-global-risks-report-2016}.
\newblock Accessed: 2018-07-31

\bibitem{erkens2012corporate}
D.H. Erkens, M.~Hung, P.~Matos, Journal of corporate finance \textbf{18}(2),
  389 (2012)

\bibitem{pielke2008normalized}
R.A. Pielke~Jr, J.~Gratz, C.W. Landsea, D.~Collins, M.A. Saunders, R.~Musulin,
  Natural Hazards Review \textbf{9}(1), 29 (2008)

\bibitem{lee2003explaining}
J.~Lee~Ray, Conflict Management and Peace Science \textbf{20}(2), 1 (2003)

\bibitem{sanchez2002soil}
P.A. Sanchez, Science \textbf{295}(5562), 2019 (2002)

\bibitem{pogge2005world}
T.~Pogge, Ethics \& international affairs \textbf{19}(1), 1 (2005)

\bibitem{rid2015attributing}
T.~Rid, B.~Buchanan, Journal of Strategic Studies \textbf{38}(1-2), 4 (2015)

\bibitem{harding2016panama}
L.~Harding, The Guardian \textbf{5} (2016)

\bibitem{lin2017limits}
X.~Lin, A.~Moussawi, G.~Korniss, J.Z. Bakdash, B.K. Szymanski, Scientific
  reports \textbf{7}(1), 6699 (2017)

\bibitem{baloi2003modelling}
D.~Baloi, A.D. Price, International journal of project management
  \textbf{21}(4), 261 (2003)

\bibitem{szymanski2015failure}
B.K. Szymanski, X.~Lin, A.~Asztalos, S.~Sreenivasan, Sci. Rep.
  \textbf{5}(10998) (2015)

\bibitem{niu2017evolution}
X.~Niu, A.~Moussawi, N.~Derzsy, X.~Lin, G.~Korniss, B.K. Szymanski, in
  \emph{International Workshop on Complex Networks and their Applications}
  (Springer, 2017), pp. 1124--1134

\bibitem{niu2018evolution}
X.~Niu, A.~Moussawi, G.~Korniss, B.K. Szymanski, Applied Network Science
  \textbf{3}(1), 24 (2018)

\bibitem{wikieventportal}
Wikipedia event portal (2019).
\newblock \urlprefix\url{https://en.wikipedia.org/wiki/Portal:Current\_events}.
\newblock Accessed: 2019-05-22

\bibitem{tran2014indexing}
G.B. Tran, M.~Alrifai, in \emph{Proceedings of the 23rd International
  Conference on World Wide Web} (ACM, 2014), pp. 511--516

\end{thebibliography}

\section*{Appendix: Tables 4-8}

\begin{table*}
\scriptsize
\centering
\begin{tabular}{|C{0.03\textwidth}|C{0.3\textwidth}|C{0.25\textwidth}|C{0.25\textwidth}|} \hline
%\begin{tabular}{| l | l | l | l |} \hline
\cellcolor{gray}ID & \cellcolor{gray}Risk & \cellcolor{gray}Tag & \cellcolor{gray}Keyword  \\ \hline
\multirow{4}{*}{1} & \multirow{4}{*}{\parbox{0.2\textwidth}{\centering\color{blue}Asset bubbles in a major economy}} & housing bubble & housing, mortgage, estate \\ \cline{3-4} 
  &  & stock bubble & share, stock, exchange \\ \cline{3-4} 
  &  & commodity bubble & commodit- \\ \cline{3-4} 
  &  & asset bubble & liquidity, asset \\ \hline
\multirow{2}{*}{2} & \multirow{2}{*}{\parbox{0.2\textwidth}{\centering\color{blue}Deflation in a major economy}} & deflation & deflation \\ \cline{3-4} 
  &  & decline in prices & price \\ \hline
\multirow{2}{*}{3} & \multirow{2}{*}{\parbox{0.3\textwidth}{\centering\color{blue}Failure of a major financial mechanism or institution}} & bank failure & bank \\ \cline{3-4} 
  &  & financial crisis & financial \\ \hline
\multirow{4}{*}{4} & \multirow{4}{*}{\parbox{0.3\textwidth}{\centering\color{blue}Failure/shortfall of critical infrastructure}} & oil/gas network failure & pipe \\ \cline{3-4} 
  &  & water supply failure & pipe \\ \cline{3-4} 
  &  & power grid failure & power, grid, electricity, blackout \\ \cline{3-4} 
  &  & civil flight failure & flight, airport, airline, plane, aircraft \\ \hline
\multirow{2}{*}{5} & \multirow{2}{*}{\parbox{0.2\textwidth}{\centering\color{blue}Fiscal crises in key economies}} & sovereign debt crisis & debt, loan, lend \\ \cline{3-4} 
  &  & liquidity crisis & liquid \\ \hline
6 & {\color{blue}High structural unemployment or underemployment} & unemployment & employ, job \\ \hline
\multirow{6}{*}{7} & \multirow{6}{*}{\parbox{0.3\textwidth}{\centering\color{blue}Illicit trade}} & drug trade & drug \\ \cline{3-4} 
  &  & smuggling & smuggl- \\ \cline{3-4} 
  &  & tax evasion & tax \\ \cline{3-4} 
  &  & money laundering & launder \\ \cline{3-4} 
  &  & human trafficking & trafficking \\ \cline{3-4} 
  &  & counterfeiting & counterfeit \\ \hline
8 & {\color{blue}Severe energy price shock} & energy price shock & price, oil, gas \\ \hline
\multirow{2}{*}{9} & \multirow{2}{*}{\parbox{0.3\textwidth}{\centering\color{blue}Unmanageable inflation}} & inflation & inflation \\ \cline{3-4} 
  &  & rising good price & price \\ \hline
\end{tabular}
\caption{Economic risks and subordinate tags. Each tag corresponds to a list of the roots of keywords.}
\label{table_risk_ids_1}
\end{table*}

\begin{table*}
\scriptsize
\centering
\begin{tabular}{|C{0.03\textwidth}|C{0.3\textwidth}|C{0.25\textwidth}|C{0.25\textwidth}|} \hline
%\begin{tabular}{| l | l | l | l |} \hline
\cellcolor{gray}ID & \cellcolor{gray}Risk & \cellcolor{gray}Tag & \cellcolor{gray}Keyword  \\ \hline
\multirow{4}{*}{10} & \multirow{4}{*}{\parbox{0.3\textwidth}{\centering\color{forestgreen}Extreme weather events}} & storm & hurricane, storm, typhoon, cyclone \\ \cline{3-4} 
  &  & blizzard & snow, blizzard, hailstone \\ \cline{3-4} 
  &  & torrential rain & rain \\ \cline{3-4} 
  &  & flood & flood \\ \hline
\multirow{6}{*}{11} & \multirow{6}{*}{\parbox{0.3\textwidth}{\centering\color{forestgreen}Failure of climate-change mitigation and adaptation}} & death & death, dead, die, toll, kill, corpse \\ \cline{3-4} 
  &  & missing & missing, disappear \\ \cline{3-4} 
  &  & injury & injur-, hurt, wound \\ \cline{3-4} 
  &  & homeless & home \\ \cline{3-4} 
  &  & damage & collaps, damage, destory \\ \cline{3-4} 
  &  & economic loss & cost, loss, dollar, usd, euro, \$, \euro \\ \hline
\multirow{2}{*}{12} & \multirow{2}{*}{\parbox{0.3\textwidth}{\centering\color{forestgreen}Major biodiversity loss and ecosystem collapse}} & ecosystem collapse & ecosystem \\ \cline{3-4} 
  &  & biodiversity loss & biodiversity, species, wildlife \\ \hline
\multirow{5}{*}{13} & \multirow{5}{*}{\parbox{0.3\textwidth}{\centering\color{forestgreen}Major natural disasters}} & earthquake & earthquake, temblor \\ \cline{3-4} 
  &  & volcanic activity & volcan- \\ \cline{3-4} 
  &  & tsunami & tsunami \\ \cline{3-4} 
  &  & landslide & slide \\ \cline{3-4} 
  &  & natural wildfire & wildfire \\ \hline
\multirow{5}{*}{14} & \multirow{5}{*}{\parbox{0.3\textwidth}{\centering\color{forestgreen}Man-made environmental damage and disasters}} & radioactive contamination & radioact- \\ \cline{3-4} 
  &  & environment contamination & contamin- \\ \cline{3-4} 
  &  & urban pollution & pollution \\ \cline{3-4} 
  &  & oil spill & oil, spill \\ \cline{3-4} 
  &  & manmade wildfire & wildfire \\ \hline
\end{tabular}
\caption{Environmental risks and and their corresponding tags. Each tag must match at least one entry in the list of the roots of keywords.}
\label{table_risk_ids_2}
\end{table*}

\begin{table*}
\scriptsize
\centering
\begin{tabular}{|C{0.03\textwidth}|C{0.3\textwidth}|C{0.25\textwidth}|C{0.25\textwidth}|} \hline
%\begin{tabular}{| l | l | l | l |} \hline
\cellcolor{gray}ID & \cellcolor{gray}Risk & \cellcolor{gray}Tag & \cellcolor{gray}Keyword  \\ \hline
15 & {\color{orange}Failure of national governance} & corruption & corruption \\ \hline
\multirow{3}{*}{16} & \multirow{3}{*}{\parbox{0.3\textwidth}{\centering\color{orange}Interstate conflict with regional consequences}} & trade war & trade, currency \\ \cline{3-4} 
  &  & military war & military, invasion, occupation, conflict \\ \hline
17 & {\color{orange}Large-scale terrorist attacks} & terrorist attack & terror \\ \hline
\multirow{2}{*}{18} & \multirow{2}{*}{\parbox{0.3\textwidth}{\centering\color{orange}State collapse or crisis}} & military coup & coup \\ \cline{3-4} 
  &  & civil war & civil, war \\ \hline
19 & {\color{orange}Weapons of mass destruction} & nuclear weapon & nuclear \\ \hline
\end{tabular}
\caption{Geopolitical risks and and their corresponding tags. Each tag must match at least one entry in the list of the roots of keywords.}
\label{table_risk_ids_3}
\end{table*}

\begin{table*}
\scriptsize
\centering
\begin{tabular}{|C{0.03\textwidth}|C{0.3\textwidth}|C{0.25\textwidth}|C{0.25\textwidth}|} \hline
%\begin{tabular}{| l | l | l | l |} \hline
\cellcolor{gray}ID & \cellcolor{gray}Risk & \cellcolor{gray}Tag & \cellcolor{gray}Keyword  \\ \hline
\multirow{3}{*}{20} & \multirow{3}{*}{\parbox{0.3\textwidth}{\centering\color{red}Failure of urban planning}} & urban pollution & pollution \\ \cline{3-4} 
  &  & factory explosion & factory \\ \cline{3-4} 
  &  & building collapse & collapse \\ \hline
\multirow{5}{*}{21} & \multirow{5}{*}{\parbox{0.3\textwidth}{\centering\color{red}Food crises}} & starvation & starv-, hunger, famin- \\ \cline{3-4} 
  &  & crop destruction & crop, agricult- \\ \cline{3-4} 
  &  & low quality of food & poison, taint, contaminat-, poultry, meat, pork \\ \cline{3-4} 
  &  & rising food price & price \\ \cline{3-4} 
  &  & food shortage & shortage, suppl- \\ \hline
\multirow{4}{*}{22} & \multirow{4}{*}{\parbox{0.3\textwidth}{\centering\color{red}Large-scale involuntary migration}} & evacuation & evacuat-, evacuee \\ \cline{3-4} 
  &  & refugee flee & refugee, flee \\ \cline{3-4} 
  &  & migration & migrant, migrate \\ \cline{3-4} 
  &  & eviction & evict \\ \hline
\multirow{5}{*}{23} & \multirow{5}{*}{\parbox{0.3\textwidth}{\centering\color{red}Profound social instability}} & riot & riot \\ \cline{3-4} 
  &  & unrest & unrest \\ \cline{3-4} 
  &  & demonstration & demonstration \\ \cline{3-4} 
  &  & march & march \\ \cline{3-4} 
  &  & revolution & revolution \\ \hline
\multirow{9}{*}{24} & \multirow{9}{*}{\parbox{0.3\textwidth}{\centering\color{red}Rapid and massive spread of infectious diseases}} & avian influenza & bird, avian, h5n1, h6n1, h7n9 \\ \cline{3-4} 
  &  & swine influenza & h1n1, swine \\ \cline{3-4} 
  &  & equine influenza & equine \\ \cline{3-4} 
  &  & aids & aids, hiv \\ \cline{3-4} 
  &  & cholera & cholera \\ \cline{3-4} 
  &  & fever & fever \\ \cline{3-4} 
  &  & diarrhea & diarrhea \\ \cline{3-4} 
  &  & ebola &ebola \\ \cline{3-4} 
  &  & epidemic & epidemic, outbreak, pandemic \\ \hline
\multirow{2}{*}{25} & \multirow{2}{*}{\parbox{0.3\textwidth}{\centering\color{red}Water crises}} & water shortage & thirst, shortage, deprivat-, suppl- \\ \cline{3-4} 
  &  & water contamination & water \\ \hline
\end{tabular}
\caption{Societal risks and and their corresponding tags. Each tag must match at least one entry in the list of the roots of keywords.}
\label{table_risk_ids_4}
\end{table*}

\begin{table*}
\scriptsize
\centering
\begin{tabular}{|C{0.03\textwidth}|C{0.3\textwidth}|C{0.25\textwidth}|C{0.25\textwidth}|} \hline
%\begin{tabular}{| l | l | l | l |} \hline
\cellcolor{gray}ID & \cellcolor{gray}Risk & \cellcolor{gray}Tag & \cellcolor{gray}Keyword  \\ \hline
\multirow{3}{*}{26} & \multirow{3}{*}{\parbox{0.3\textwidth}{\centering\color{indigo}Adverse consequences of technological advances}} & genetically modified food & genetic \\ \cline{3-4} 
  &  & radioactive contamination & radioact- \\ \cline{3-4} 
  &  & pyrotechnics & pyrotech- \\ \hline
\multirow{2}{*}{27} & \multirow{2}{*}{\parbox{0.3\textwidth}{\centering\color{indigo}Breakdown of critical information infrastructure}} & internet blackout & internet \\ \cline{3-4} 
  &  & satellite break & satellite \\ \hline
\multirow{8}{*}{28} & \multirow{8}{*}{\parbox{0.3\textwidth}{\centering\color{indigo}Large-scale cyberattacks}} & cyberattack & cyber \\ \cline{3-4} 
  &  & malware & malware \\ \cline{3-4} 
  &  & computer worm & worm \\ \cline{3-4} 
  &  & software virus & virus \\ \cline{3-4} 
  &  & trojan horse & trojan \\ \cline{3-4} 
  &  & hacking attack & hack \\ \cline{3-4} 
  &  & DOS attack & denial \\ \cline{3-4} 
  &  & online attack & online \\ \hline
\multirow{2}{*}{29} & \multirow{2}{*}{\parbox{0.3\textwidth}{\centering\color{indigo}Massive incident of data fraud/theft}} & data leak & leak, classified, secret \\ \cline{3-4} 
  &  & hacking data & hack \\ \hline
\end{tabular}
\caption{Technological risks and and their corresponding tags. Each tag must match at least one entry in the list of the roots of keywords.}
\label{table_risk_ids_5}
\end{table*}

%\begin{acknowledgements}
%\end{acknowledgements}




\end{document}